\documentclass[twocolumn,preprintnumbers, amsmath,amssymb,pof]{revtex4-1}

\usepackage{graphicx}
\usepackage{dcolumn}
\usepackage{bm}
\usepackage[french]{babel}
\usepackage[french]{babel}
\usepackage[latin1]{inputenc}
\usepackage{amsmath}
\usepackage{amsfonts}
\usepackage{amssymb}
\usepackage{graphicx}
\usepackage{subfigure}
\usepackage{epsfig}
\usepackage{array}
\usepackage{enumerate}
\usepackage{indentfirst}
\usepackage[T1]{fontenc}
\usepackage[french]{varioref}
\usepackage{float}
\usepackage{subeqnarray}
\usepackage{units}
\usepackage{textcomp}
\usepackage{layout}
\usepackage{picins}
\usepackage{xcolor}

\def\mcm{\nobreak\mbox{$\;$\textnormal{\textmu m}}}

\def\mNm{\nobreak\mbox{$\;$\textnormal{mN\,m}$^{-1}$}}
\def\kgm{\nobreak\mbox{$\;$\textnormal{kg\,m}$^{-3}$}}

\def\degC{\nobreak\mbox{$\;$\textnormal{°C}}}

\def\Pas{\nobreak\mbox{$\;$\textnormal{Pa.s}}}

\def\nm{\nobreak\mbox{$\;$\textnormal{nm}}}

\newcommand\al{\textit{et~al.}\ }

\newcommand{\purple}[1]{\textcolor{violet}{#1}}
\newcommand{\cyan}[1]{\textcolor{cyan}{#1}}
\newcommand{\blue}[1]{\textcolor{blue}{#1}}
\newcommand{\green}[1]{\textcolor{green}{#1}}
\newcommand{\orange}[1]{\textcolor{orange}{#1}}
\newcommand{\red}[1]{\textcolor{red}{#1}}
\newcommand{\gray}[1]{\textcolor{gray}{#1}}

\begin{document}

\title{Accelerated drop detachment in granular suspensions}

\author{Claire Bonnoit}
\author{Thibault Bertrand}
\author{Eric Cl\'{e}ment}

\author{Anke Lindner\footnote{corresponding author: anke.lindner@espci.fr}}%

\affiliation{{Laboratoire de Physique et M\'ecanique des Milieux
H\'et\'erog\`enes (PMMH)\\
UMR 7636 CNRS - ESPCI - Universit\'es
Paris
6 et 7, \\
10, rue Vauquelin, 75231 Paris Cedex 05, France }}%

\date{\today}

\begin{abstract}

We experimentally study the detachment of drops of granular
suspensions using a density matched model suspension with varying
volume fraction ($\phi=15\%$ to $55\%$) and grain diameter
($d=20\mcm$ to $140\mcm$). We show that at the beginning of the
detachment process, the suspensions behave as an effective fluid.
The detachment dynamics in this regime can be entirely described by
the shear viscosity of the suspension. At later stages of the detachment the
dynamics become independent of the volume fraction and are found
to be identical to the dynamics of the interstitial fluid.
Surprisingly, visual observation reveals that at this stage
particles are still present in the neck. We suspect rearrangements
of particles to locally free the neck of grains, causing the observed
dynamics. Close to the final pinch off, the
detachment of the suspensions is further accelerated, compared to
the dynamics of pure interstitial fluid. This acceleration might be
due to the fact that the neck diameter gets of the order of magnitude of the size of the grains and a continuous thinning of the liquid thread is not possible any more. The crossover between the different detachment regimes is function of the grain size
and the initial volume fraction. We characterize the overall acceleration as a function of the grain
size and volume fraction.

\end{abstract}

\maketitle

\section{Introduction}
\label{sec:Introduction}

The stability of jets or the detachment of drops is an important
issue in everyday life as well as for industrial applications. The
formation of drops can be important for simple liquids, complex
fluids or even dense suspensions or pastes. Especially in the food
industry concentrated suspensions or pastes are processed, extruded
from nozzles or filled in recipients. The formation of drops of
liquids laden with particles is of great importance for example for
inkjet printing  \cite{Morrison2010}.

The break-up of liquid jets of simple liquids into drops was first
described by Plateau and Rayleigh \cite{Plat73, Rayl79} in the
nineteenth century. The final separation of two drops, the so called
"pinch-off" represents a finite time singularity and has attracted
lot of interest lately. Many experimental or theoretical studies
report the description of the non-linear behavior of drop
separation, reviewed by Eggers \cite{Egge97}.

The fact that the dynamics of the detachment are now well known for
simple liquids has also initiated the use of capillary breakup as a
rheological technique, allowing to directly access the viscosity of
a given fluid \cite{McKinley2000}. The thinning of the filament,
leading to the "pinch-off" represents a strong elongation and thus
allows to study viscous properties under elongation. Experiments
with polymer solutions have shown that this experimental situation
is indeed a good set-up to access elongational properties of these
fluids \cite{Amarouchene2001, Tirtaatmadja2006, Sattler2008}.

The stability of jets formed by foams, pastes, suspensions, dry grains, or grains
immersed in a surrounding liquid has also been addressed
\cite{Pignatel2009, Lohse2004, Royer2009, Lespiat2010, Coussot2005, Furb04,Furb07}.

Furbank \al  \cite{Furb04,Furb07} have studied drop thread dynamics of particle-laden liquids. They have shown that at early stages of the detachment the suspension can be described as an effective fluid. They have also observed that this regime is followed by a second regime, dependent on the characteristics of the individual particles. In this paper we present a detailed study of the later stages of the detachment process, extending in this way the study of Furbank \al  \cite{Furb04,Furb07}. Comparing to
pure oils matching either the shear viscosity of the suspensions \cite{Bonnoit2010, Bonnoit_PRL_2010} or the interstitial fluid we
show that the detachment takes place through different regimes. We study the cross over between these regimes as a function of the grain diameter and the volume fraction.  We show that the detachment of drops of granular suspensions is accelerated, not only compared to a
simple fluid, matching the shear viscosity of the suspensions but
also, at later stages, compared to the dynamics of the interstitial fluid. The overall acceleration
is quantified as a function of the volume fraction and the particle
diameter.

The paper is organized as follows. In section
\ref{sec:Theory} we recall the theory of drop detachment of
viscous Newtonian fluids. Section \ref{sec:exp set_up} describes the
experimental set-up and the model system. We then introduce the different detachment regimes (section \ref{sec:detachment regimes})
and discuss the transition between these regimes in section \ref{sec:crossover detachment}. The acceleration is quantified in  section \ref{sec:accelerated detachment} and we conclude in section \ref{sec:conclusion}.

\section{Theory: detachment of a Newtonian fluid}
\label{sec:Theory}


A pending drop starts to fall, when gravitational forces overcome
surface tension. During detachment of the drop surface tension acts
as the main pinching force and tends to thin the filament linking
the drop to fluid left at the nozzle and gravity becomes negligible.
The dynamics of the thinning of the thread are given by a
competition between capillarity that tends to thin the filament and
viscosity and/or inertia that resist the thinning of the thread.

Such a detachment of a Newtonian fluid can be divided into different
regimes. The first regime is governed by linear instability
 described in the studies on the stability of liquid jets by Plateau
 and Rayleigh \cite{Plat73,Rayl79}. It corresponds to the instability of a
 liquid cylinder disintegrating into individual drops. The exponential growth
 of the most unstable mode leads to an exponential evolution of the
 minimal diameter as a function of time with a given growth rate,
 depending on the material parameters of the system. In this regime
 the evolution of the detachment is dependent on the initial experimental conditions.

 Close to the final detachment,
 the equations of motion become non linear \cite{Egge97}.
 The "pinch-off" is a finite time singularity, localized in space and time. The dynamics thus become
 independent of the initial conditions or from the particular set-up used.
 The motion of the fluid can be reduced to a one dimensional problem and the
 profile of the drop is then governed by universal scaling
 laws in a self-similar flow \cite{Pere90,Egge97}.

Considering a viscous liquid Papageorgiou \cite{Papa95} describes
theoretically the thinning dynamics of the filament resulting of a competition between capillarity and viscosity, called
the Stokes regime. More precisely, the minimum diameter $W$ of the
filament (see figure \ref{fig:setup}) can be written as:
\begin{equation}\label{eq:papageo}
   W(t)= 2 \times 0.0708 \times \frac{\gamma}{\eta} (t_{\mathrm{p}}-t) \ ,
\end{equation}
with $t_{\mathrm{p}}$ the time of pinch-off, $\gamma$ the surface
tension and $\eta$ the viscosity of the fluid.

Close to the pinch-off, the radius of curvature tends to zero,
leading to an infinite pressure. Then the velocity of the motion
diverges and inertia can not be neglected any more. Eggers gives a
theoretical description of this regime of final detachment
\cite{Egge93}, called the Navier-Stokes regime which results of a
competition of capillarity, viscosity and inertia. The minimum
diameter $W$ can thus be written as a linear scaling:
\begin{equation}\label{eq:eggers}
   W(t)= 2 \times 0.0304 \times \frac{\gamma}{\eta} (t_{\mathrm{p}}-t) \ ,
\end{equation}
Experiments have prooven the existence of those three different
regimes \cite{Kowa96,Roth03} for a viscous fluid.

In summary the drop detachment for a pure viscous fluid takes place via
three different regimes, characterized by an exponential regime
followed by two linear regimes of different slope. Usually one
chooses the final pinch off as the reference in time, but strictly
speaking this is relevant only for the last regime before final
detachment and there is no absolute reference on time for the other
detachment regimes. This becomes even more important for
non-Newtonian fluids, where different detachment dynamics can occur
as a function of the nature of the fluid. To be able to compare the
dynamics far away from the final pinch it might be necessary to use
another reference in time instead of $t_p$. This is commonly done
for example for the study of the detachment of complex fluids \cite{Tirtaatmadja2006}
and is also what we will do in our analysis.


\section{Experimental set-up and model system}
\label{sec:exp set_up}


\subsection{Model suspension and viscous oils}
\label{subsubsec:suspensions}

The model suspensions are formed by spherical polystyrene beads from
Dynoseeds with different grain diameters $d=20, 40, 80, 140 \mcm$
and density $\rho=1050-1060 \kgm$ . The grains almost have a perfect
sphericity as guaranteed by the supplier (the ratio of the standard
deviation to the mean diameter of the grains is less then $5\%$). The roughness
of the beads has been determined by AFM measurement to be of $100\nm$ by Deboeuf
\al \cite{Debo09}. The grains are dispersed in a silicon oil
(Shin Etsu SE KF-6011.) We measured the viscosity of the pure oil
$\eta_0=0.18\Pas$ and its surface tension $\gamma =21 \pm 1
\mNm$ at
$T=21^\circ$C. Its density is $\rho=1070\kgm$ as given by the supplier. We vary the
volume fraction $\phi=V_{\mathrm{g}}/V_{0}$ defined as the volume of
grains $V_{\mathrm{g}}$ on the total volume $V_{0}$ from $15\%$ to
$55\%$. In this way we prepare monodisperse non-buoyant model suspensions where Brownian motion can be neglected.

\begin{figure}[h]
\begin{center}
 \includegraphics[width=8cm]{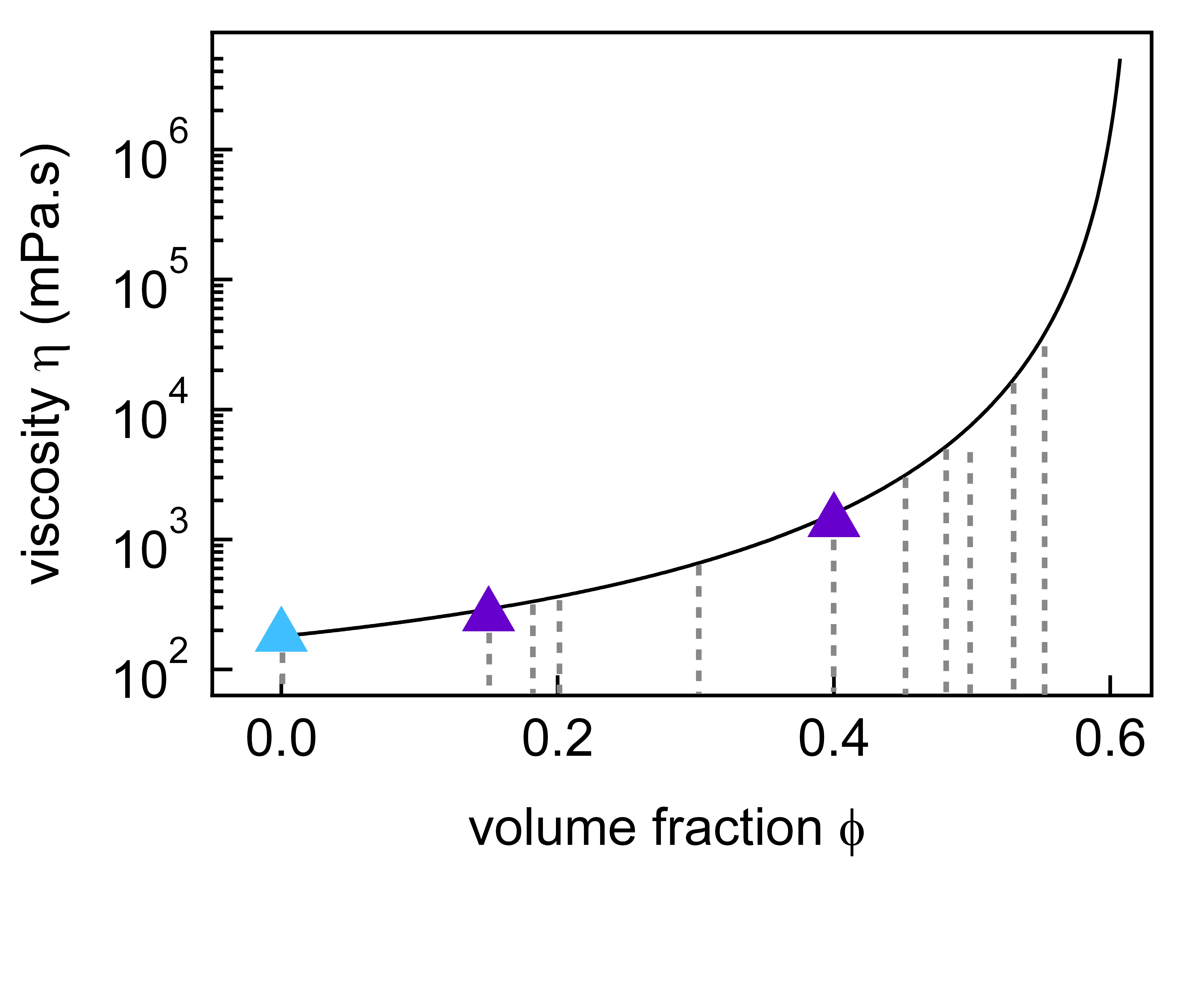}\\
 \end{center}
  \caption{Viscosities $\eta$ as a function of the volume fraction $\phi$. The solid line represents the prediction by the Zarraga model and the dashed lines indicate the volume fractions of the suspensions used in this study ($\phi=15$,~$18$,~$20$,~$30$,~$40$,~$45$,~$48$,~$50$,~$ 53$,~$ 55\%$). Triangles 
  represent the pure oils used (\cyan{cyan}
   for the interstitial oil and \purple{purple}
   for the pure oils approximatively matching the shear viscosity of the suspensions of $\phi=15\%$ and $\phi=40\%$).}
  \label{fig:rheol}
\end{figure}

In a previous study \cite{Bonnoit2010, Bonnoit_PRL_2010} we have characterized the
shear viscosity of this model suspension combining classical and
inclined plane rheometry and we have shown that the shear viscosity
$\eta (\phi)$ of the suspensions is well described by the Zarraga
model \cite{Zarraga_hill}:
\begin{equation}
    \eta_{\mathrm{Z}}(\phi)=\eta_0\frac{\exp(-2.34\phi)}{(1-\phi/\phi_{\mathrm{m}})^{3}}, {\quad
    \phi_{\mathrm{m}}\approx0.62}
    \label{equation:Zarraga}
\end{equation}
where $\eta_0$ is the viscosity of the interstitial fluid. We use this model to estimate the viscosities of the suspensions used in the present paper (see figure \ref{equation:Zarraga}). We also use pure oils AP200 and AP1000 from Sigma Aldrich matching approximately the viscosities of the suspensions of $\phi=15\%$ and $\phi=40\%$. Their viscosities measured at $T=21^\circ$C are given in table \ref{table:viscosities} together with the viscosities of the suspensions. The error in the suspension viscosity is due to an uncertainty of $\phi\pm 1\%$. The surface tension of the pure oils is $20\mNm$ at $T=21^\circ$C.

\begin{table}[h]
\begin{center}
    \begin{tabular}{|c|c||c|c|}
    \hline
        \multicolumn{1}{|c|}{Suspensions} &
        \multicolumn{1}{c||}{ $\eta_{Susp}$ ($Pa.s$)} & \multicolumn{1}{|c|}{Pure oils} & \multicolumn{1}{c|}{ $\eta_{PO}$ ($Pa.s$)}\\ \hline
        &&Shin Etsu ($\eta_0$)& 0.18\\
       $ \phi=15\%$ & $0.29 \pm 0.01$ &  AP200& 0.26\\
        $\phi=40\% $& $1.55 \pm 0.1$ &  AP1000& 1.4\\ \hline
    \end{tabular}
\caption{\label{table:viscosities} Table of the viscosities. The viscosities of the suspensions are calculated from the Zarraga model and the viscosities of the pure oils have been measured by classical rheometry. All viscosities are given at $21\degC$}
\end{center}
\end{table}

\subsection{\label{subsec:appa}Experimental set-up and method}

\begin{figure}[htb]
     \centering
    \includegraphics[width=0.49\textwidth]{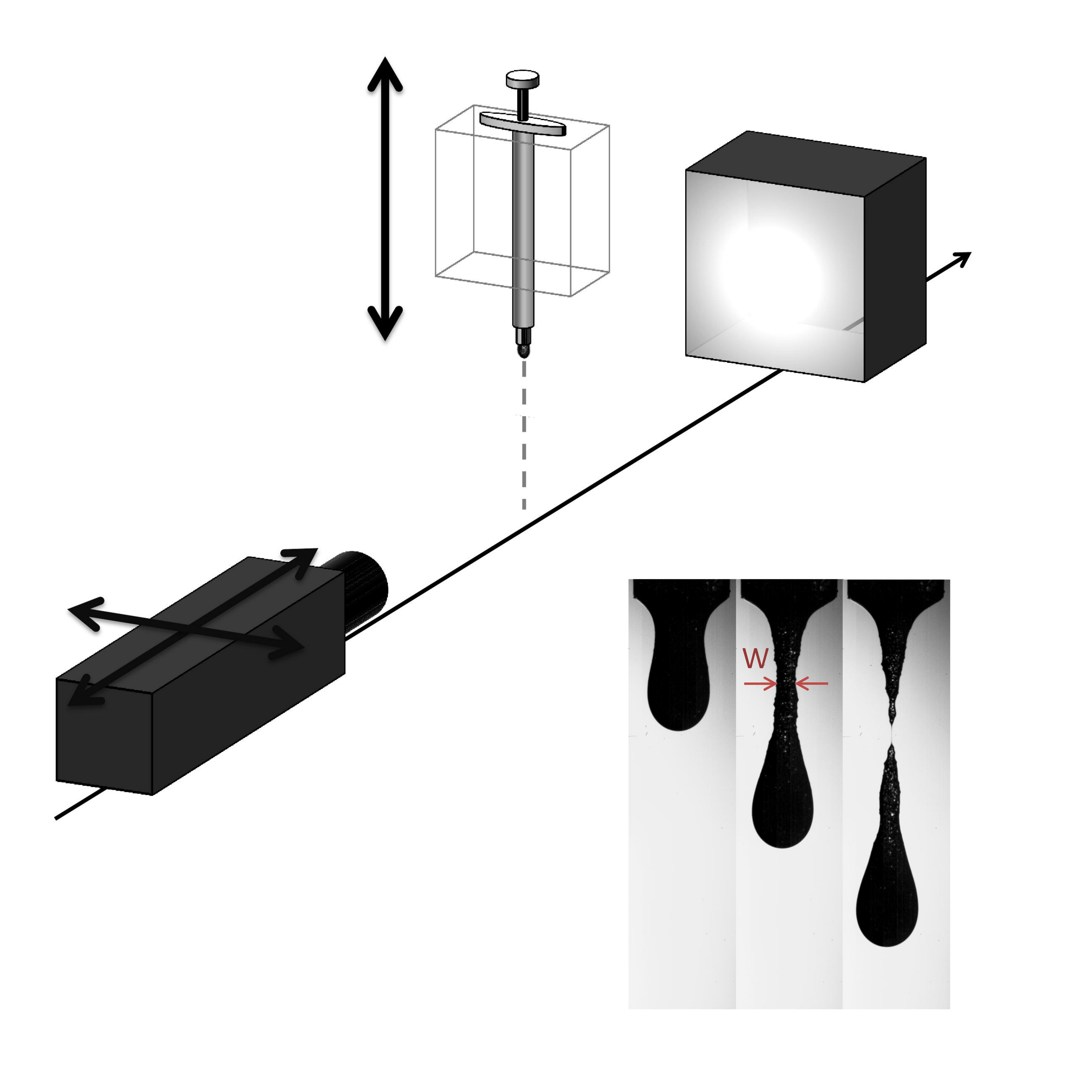}
     \caption{ Schematic of the experimental set-up. The suspension is extruded from a syringe of external diameter of 4~mm. The subsequent drop formation is observed via a fast camera. Backlight illumination provides good quality of the pictures. We measure the minimum neck diameter $W$ from the pictures as a function of time $t$. The last snapshot represents the moment of final pinch off $t_p$.}
     \label{fig:setup}
\end{figure}

All experiments are performed with the experimental set-up presented
on figure \ref{fig:setup}. The suspension is filled in a syringe and manually extruded from the latter.
The inner diameter of the nozzle is 1.5 mm and the outer diameter is
4 mm. The nozzle is wetted by the fluid and the size of the drop
corresponds in the beginning to the outer diameter of the nozzle.
The syringe and the camera are mounted on translation stages to
place the final detachment in the window of observation of the
camera.

A fast camera (Photron Fastcam SA3) records the detachment, with a
frame rate of $4\,000$ frames per second at a resolution of 1024*256
pixels. We use a macro lens (Sigma 105 mm F 2.8 EX DG) in order to
zoom on the drop. Background illumination is provided by a
high-brightness LED multihead lamp (300 W). We use a diffuser
between the camera and the syringe to get a homogenous illumination
of the drop.

For all experiments we determine the minimal neck diameter $W$ (see
figure \ref{fig:setup}) as a function of time $t$ during detachment.
The moment of final detachment is called $t_p$ and is used as a
temporal reference. Image processing with ImageJ is used to extract
the profiles of the drops and the minimal neck diameter. We have estimated the resolution of our data to be
of $60\mcm$ for the suspensions and of $80\mcm$ for the pure oils. This corresponds to two pixels on the
pictures. Data points below this resolution is shown in the graphs as the small grey dotted lines. The slight difference in resolution is due to the fact that we use a slightly different zoom for the pure oils compared to the suspensions. We conducted three experiments for each condition. For the pure oils the reproducibility is superior to our spatial resolution for neck diameters
smaller than $1.5$~mm. Due to the fact that we manually extrude the droplet from the syringe, the data is less reproducible at earlier stages of the detachment process. As we study the detachment close to the final pinch off this does not influence our measurements. The experiments for the suspensions are slightly less reproducible than the experiments for the pure oils. In the graphs showing the time evolution of the experiments only one experiment is represented per condition, but measurements are always done for three experimental realizations, reflecting in this way the reproducibility.

A first set of experiments is done for a fixed grain size $d=40\mcm$
at different volume fractions $\phi=15$,~$18$,~$30$,~$40$,~$45$,~$ 48$,~$ 50$,~$ 53$,~$ 55\%$. Up to
volume fractions of $55\%$ no clogging or filtration \cite{Kulkarni2010} of the
suspension is observed at the outlet of the syringe and the
extracted volume fraction is homogeneous and reproducible. A second
set of experiments is done for suspensions with four different grain
diameters $d=20, 40, 80, 140\mcm$ and two different volume
fractions $\phi=15$,~$ 40 \%$.

The pure oils given in table \ref{table:viscosities} are equally tested. To get insight into the thinning dynamics of the suspensions we compare the results obtained for the suspensions directly to the results obtained for the pure oils, rather then fitting these results to the predictions for the different detachment regimes given in section \ref{sec:Theory}. Our approach has the advantage that we do not need to determine previously in which stage of the detachment process we are and thus allows for an unambiguous discussion of the thinning dynamics of the suspensions.


\section{\label{sec:detachment regimes} Different detachment regimes}


\begin{figure}[th!]
    \centering
    \includegraphics[width=0.49\textwidth]{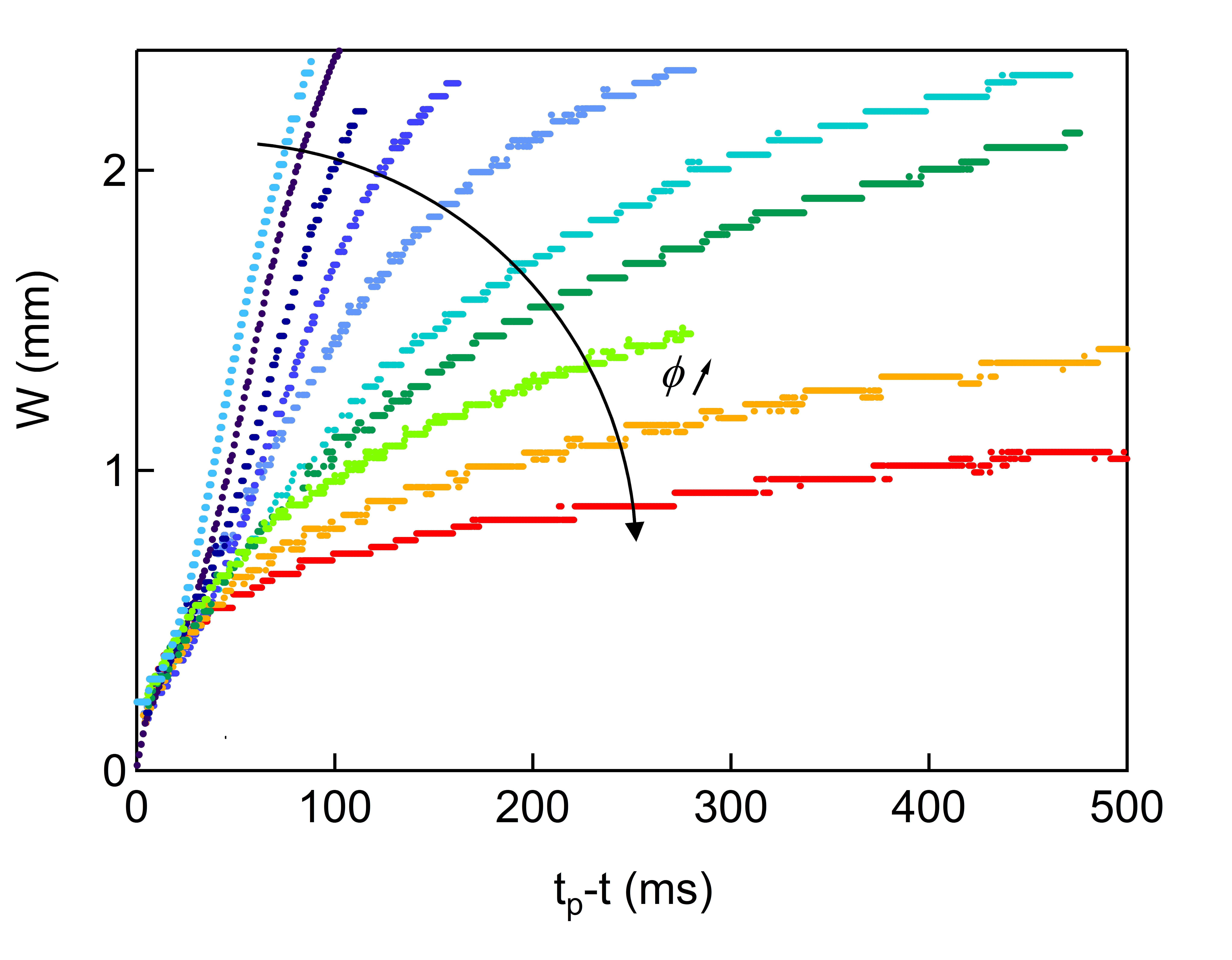}
    \caption{Time evolution of the minimum neck diameter $W$ for suspensions
    of volume fractions $\phi=15$,~$20$,~$30$,~$40$,~$45$,~$48$,~$50$,~$ 53$,~$ 55\%$
    with grain diameter $d=40\mcm$. The most left curve  (\cyan{cyan}) corresponds to pure interstitial fluid ($\phi=0$. The origin of the $x$-axis is given by the time of the pinch for each suspension, the curve for the interstitial fluid is shifted by $\Delta t_{SE}=-20$ ms.}
    \label{fig:Graph_tt_frac}
\end{figure}

The minimal neck diameter $W$ as a function of $t_p-t$ for the first set of experiments for a grain diameter of
$d=40\mcm$ and volume fractions from $\phi=15$ to $ 55\%$ are
shown on figure \ref{fig:Graph_tt_frac}. In addition the results for
the interstitial fluid ($\phi=0$) are represented.  Note that when representing the data as a function of $t_p-t$ the time goes from right to left; from the early stages of the detachment on the right side of the graph to the final pinch off on the left side. At the beginning of the experiment the evolution of
the minimum diameter depends on the initial volume fraction $\phi$.
One observes that the more dense and thus the more viscous the suspensions are,
the slower are the dynamics. The thinning of the filament takes longer for more dense suspensions.
This is in agreement with observations on viscous oils
\cite{Roth03}. At a later stage of the detachment process all the
data collapse onto a single curve. In this regime, the detachment
process is independent of the initial volume fraction
$\phi$ and is observed to be identical to the dynamics of the
interstitial fluid.

\begin{figure}[h!]
    \centering
         \includegraphics[width=0.35\textwidth]{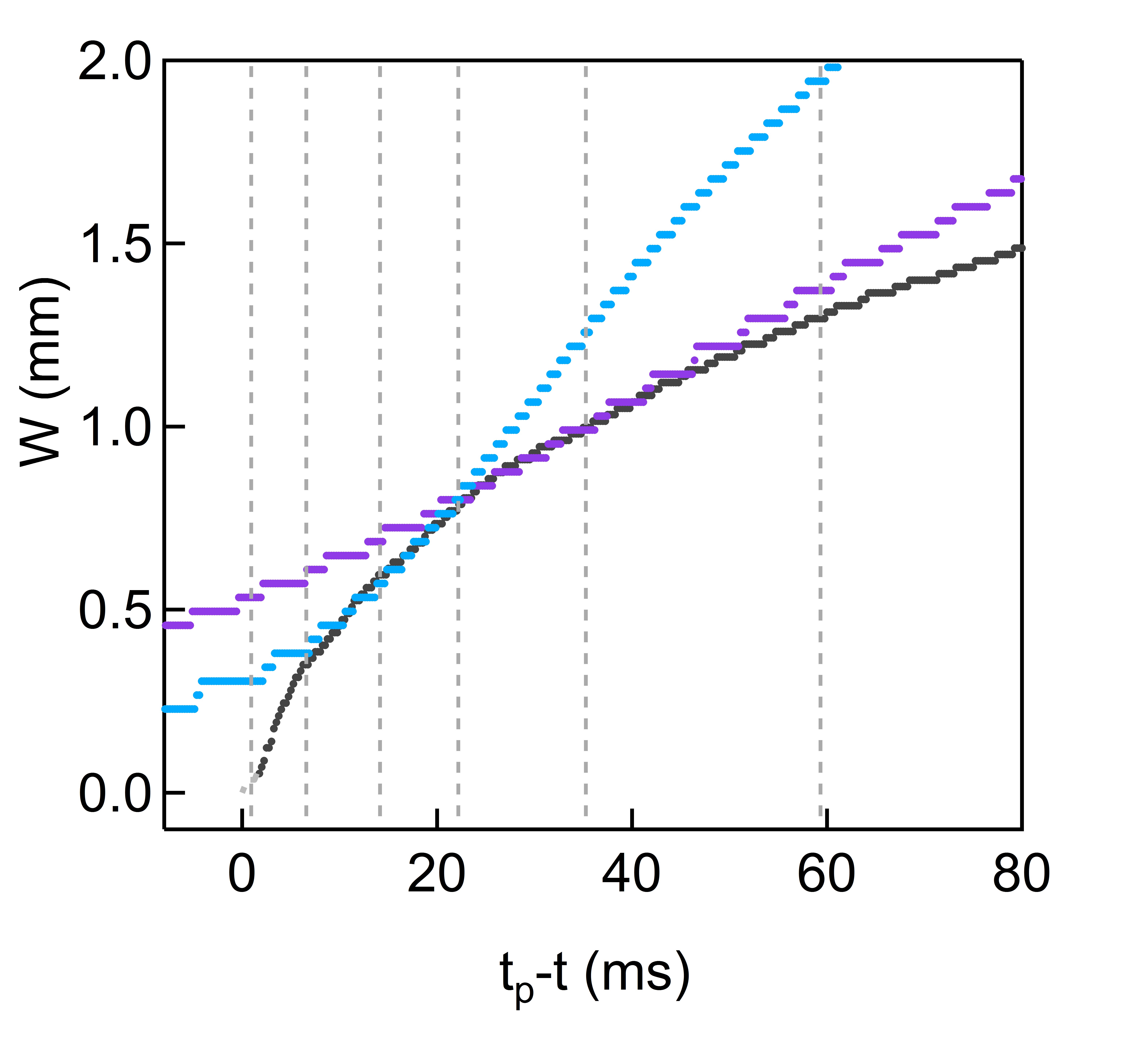}\\
          \includegraphics[width=0.3\textwidth]{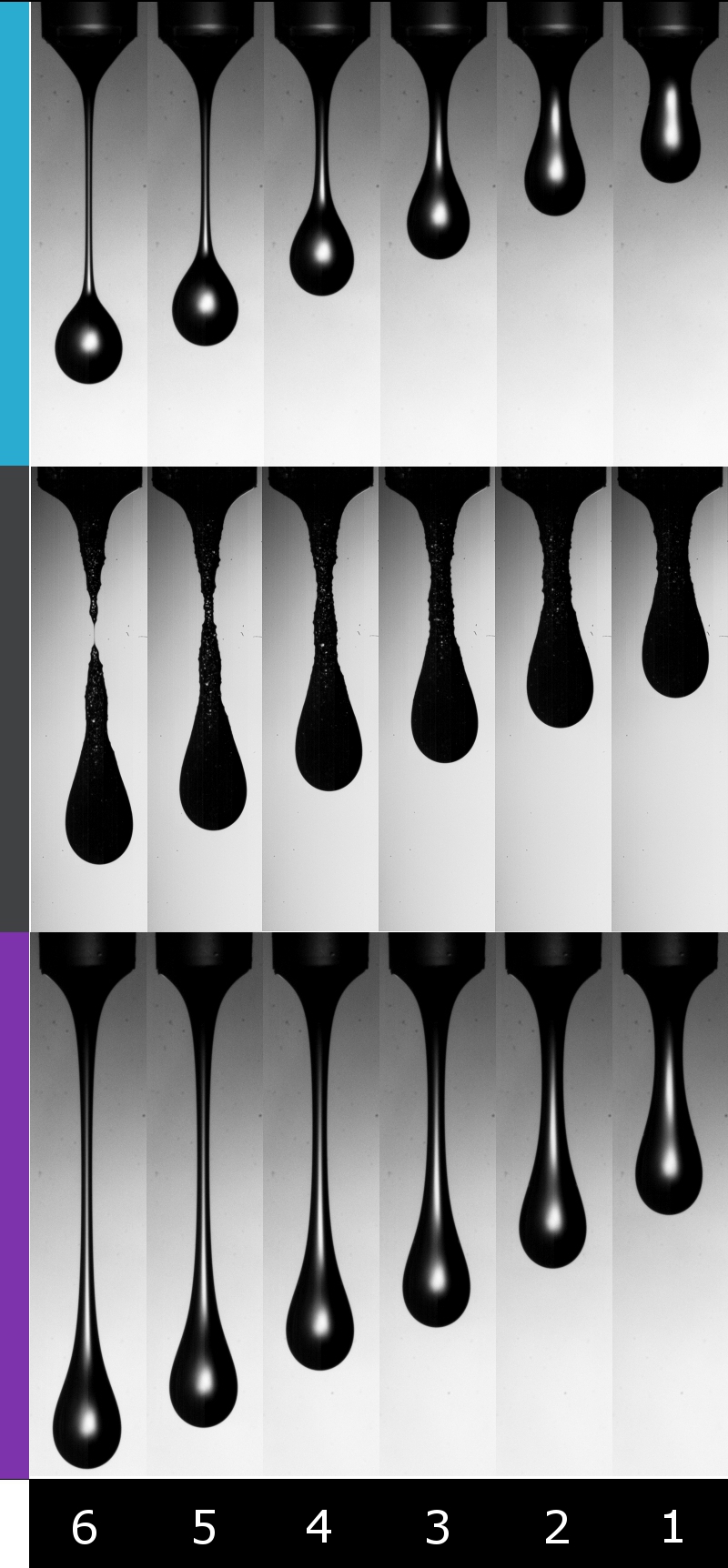}
    \caption{(\textit{top}) Evolution of the minimal neck diameter for a suspensions with $d=140µm$ and $\phi=40\%$ (\gray{gray}), for the pure interstitial oil Shin Etsu (\cyan{cyan}) and for the pure oil AP1000 (\purple{purple}) matching the shear viscosity of the suspension. The origin of the $x$-axis is given by the time of the pinch off for the suspension; the graphs for the pure oils Shin Etsu and AP1000 are shifted $\Delta t_{SE}=-30.75$ ms \& $\Delta t_{AP_{1000}}=-124$ ms respectively. (\textit{bottom}) Corresponding snapshots for the interstitial oil Shin Etsu (top), the suspension (middle) and the pure oil AP1000 (bottom). The snapshots are taken at the moments of the detachment process indicated by the dashed lines.}
    \label{fig:regimes}
\end{figure}

These observations indicate that the detachment takes place via
different regimes, a first regime that depends on the volume
fraction and a second regime where the dynamics become independent of
the volume fraction and is identical to the dynamics of the
interstitial fluid, representing the limit $\phi$ going to 0.

The results for the interstitial fluid
had to be shifted by an offset $\Delta t_{SE}=20 \mathrm{ms}$ (see section \ref{sec:Theory}) to obtain data collapse
with the suspensions onto a single curve. The pinch off for the pure oil is in this representation at $t=-20\mathrm{ms}$, not shown on figure \ref{fig:Graph_tt_frac}. This indicates that the final detachment for
the suspensions is accelerated compared to the interstitial fluid corresponding to a third detachment regime.

The different regimes can be seen in more detail on figure
\ref{fig:regimes} \textit{(top)} where $W$ is represented as a function of
$t_p-t$ for a suspension of $d=140\mcm$ and $\phi=40\%$. Results for the
pure oil AP1000 matching the effective viscosity of the suspension and the
interstitial fluid Shin Etsu are also represented. Note that the curves of the pure
oils have been shifted in time, as indicated in the caption of the figure.

From this figure it is obvious that
the first part of the detachment regime corresponds to a regime
where the dynamics are identical to an effective fluid of the
corresponding shear viscosity. We will show in the following that this also holds for other grain diameters and volume fractions (section \ref{sec:crossover detachment}). The fact that the thinning of the thread is identical for the
suspension and the corresponding pure oil shows that there is a
regime where the dynamics in an elongational flow of a granular
suspension are solely given by the effective shear viscosity of the
suspension. This indicates that the elongational
viscosity $\eta_e$ of model suspensions is solely given by the shear
viscosity $\eta$, as for Newtonian fluids where $\eta_e=3 \eta$.
This observation is in agreement with the results
by Furbank {\it et al.} \cite{Furb07} obtained using a different
method. These results are also in agreement with results by Clasen
{\it et al.} \cite{Clasen2010} on colloidal suspensions, showing
that the elongational viscosity is solely given by the shear
viscosity.

At a certain neck diameter, one
observes the crossover to the regime where the thinning dynamics are
identical to the interstitial fluid. And finally the suspensions
detach more quickly than the interstitial fluid.

In summary, we observe three different detachment regimes: I) a regime where the dynamics are governed by the shear viscosity of the suspensions, the suspensions behave as an effective fluid; II) a regime where the dynamics are given by the interstitial fluid and  III) a final detachment regime close to the final pinch off.

Figure \ref{fig:regimes} \textit{(bottom)} shows snapshots of these different regimes for the suspension and the two pure oils at the moments of the detachment process indicated by the dashed lines on the figure. On the first two snapshots (1 and 2) one observes the thinning of the neck in the effective fluid regime. The neck radius is identical between the pure oil AP1000 and the suspension. The third snapshot shows the transition towards the interstitial fluid regime. No obvious change in the form of the drops can be observed at this stage. Surprisingly one observes from these pictures that in the interstitial fluid regime (picture 4) there are still grains present in the neck. The surface of the drop of the granular suspensions starts to become more rough compared to the pure oils in this regime. In the two last pictures (5 and 6) one observes the cross over to the final detachment regime for the suspension. The suspension detaches very quickly compared to the drop of interstitial fluid that shows a slow thinning of a stable long viscous thread. It can be seen from these pictures that the deformation of the neck becomes very localized for the suspension and that the volume of the suspension that is actually deformed is reduced in this regime.


\section{\label{sec:crossover detachment}Crossover between different regimes}


In the following we discuss the crossover between the three different detachment regimes. To do so, we focuss on two selected volume fractions: $\phi=15$ and $40\%$ and vary the grain diameter from $d=20\mcm$ to $d=140\mcm$.

\subsection{\label{subsec:crossover_effective} Crossover from effective fluid to interstitial fluid regime}

                \begin{figure}[h!]%
                \centering
                \includegraphics[width=0.35\textwidth]{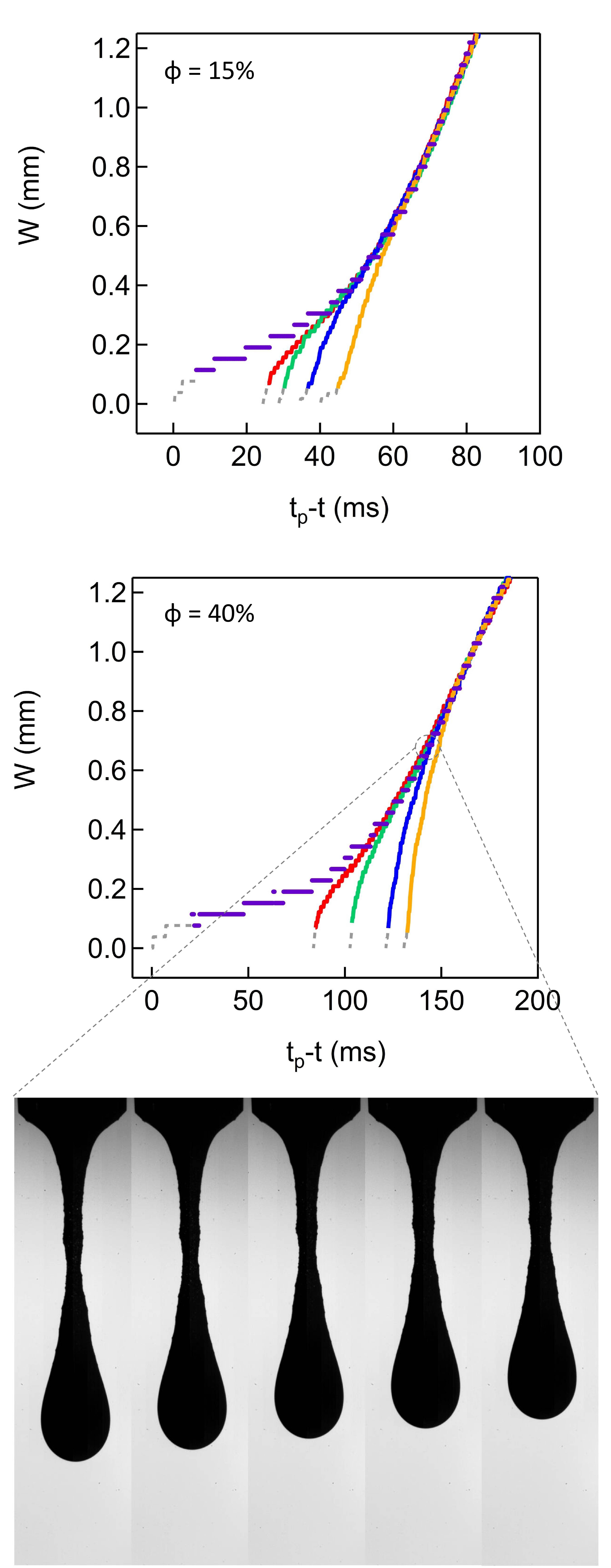}
                \caption[A set of four subfigures.]{\textit{(top)} Thinning dynamics for suspensions close to the deviation from the effective fluid regime for particle diameters of 20 $µm$ (\red{red}), 40 $µm$ (\green{green}), 80 $µm$ (\blue{blue}) and 140 $µm$ (\orange{orange}) for volume fractions of  $\phi=15\%$ and $\phi=40\%$ The (\purple{purple}) curves represent the pure oils matching the shear viscosities of the suspensions, AP200 and AP1000, respectively. The origin of the $x$-axis is given by the time of the pinch for the pure oils, the curves for the suspensions have been shifted to match the effective fluid regime ($\phi=15\%$: $\Delta t_{20}=24.50$ ms, $\Delta t_{40}=28.75$ ms, $\Delta t_{80}=34.50$ ms \& $\Delta t_{140}=40.25$ ms. $\phi=40\%$: $\Delta t_{20}=83.75$ ms, $\Delta t_{40}=102.75$ ms, $\Delta t_{80}=121.25$ ms \& $\Delta t_{140}=130.50$ ms).\textit{(bottom)} Corresponding snapshots for suspensions of $d=80 \mathrm{µm}$ particles for $\phi = 40\%$ around the transition, $\Delta t = \{-5, -2.5, 0, 2.5, 5 \}$ ms}.
                \label{fig:PO}%
                \end{figure}

Figure \ref{fig:PO} \textit{(top)} represents the minimal neck diameter for suspensions of different grain diameter and $\phi=15\%$ together with the pure oil AP200 and $\phi=40\%$ together with the pure oil AP1000, respectively. The results for the suspensions have been shifted such as to obtain the best agreement with the pure oils. The values for the corresponding offsets in time $\Delta t$ are given in the caption. For both volume fractions one observes clearly that in a certain range of $W$ we obtains very good agreement between the thinning dynamics of the suspensions and the pure oils. This shows once again that at early stages of the detachment process the dynamics are governed by the shear viscosity and the suspensions behave as an effective fluid. For neck diameters smaller than a given value of $W$, the neck thins more quickly for the suspensions than for the pure oil. This change in dynamics takes place at different values of the neck diameter that we call $W_{eff}$ and is function of the grain diameter $d$. From the graphs one observes that for the larger $d$ this change in dynamics occurs for larger $W_{eff}$.

\begin{figure}[h]
     \centering
   \includegraphics[width=0.45\textwidth]{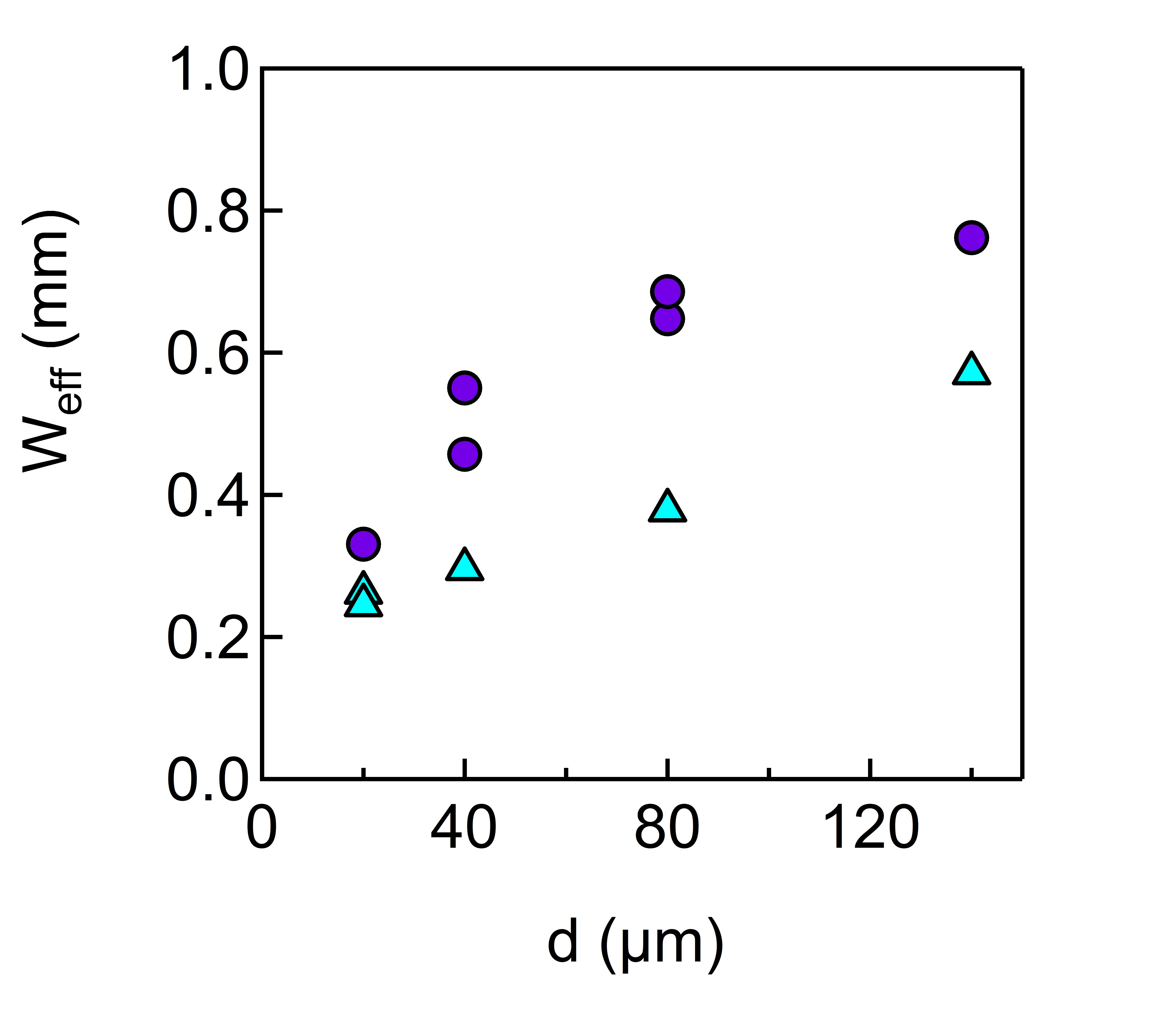}
\caption{Diameter of the neck at the crossover between the effective fluid and the interstitial fluid regime as a function of grain size for $\phi$: 15\% (\cyan{cyan}) and 40\% (\purple{purple}).}
    \label{fig:w_eff_d}
\end{figure}

We have measured $W_{eff}$ for the three runs conducted for each experimental condition. To determine $W_{eff}$ we use as a criterion that the curves for the minimal neck diameter between the suspensions and the pure oil differ by more than our experimental resolution. On figure \ref{fig:w_eff_d} $W_{eff}$ is represented  as a function of the grain diameter $d$ for the two volume fractions. $W_{eff}$ is typically of the order of 10 grains. The crossover to the interstitial fluid regime takes place at earlier stages of the detachment process (and thus at larger $W_{eff}$) for larger grain diameters. The increase in $W_{eff}$ with $d$ is observed to be less than linear. For the higher volume fraction deviations from the effective fluid regime takes place at larger neck diameters compared to the smaller volume fraction.

The snapshots shown on figure \ref{fig:PO} \textit{(bottom)} show pictures around the transition towards the interstitial fluid regime for $d=80\mcm$. They will be discussed in more detail in subsection \ref{subsection:discussion}.

\subsection{\label{subsec:crossover_SE} Crossover from the interstitial fluid to the final detachment regime}

\begin{figure}[h!]%
                \centering
                   \includegraphics[width=0.35\textwidth]{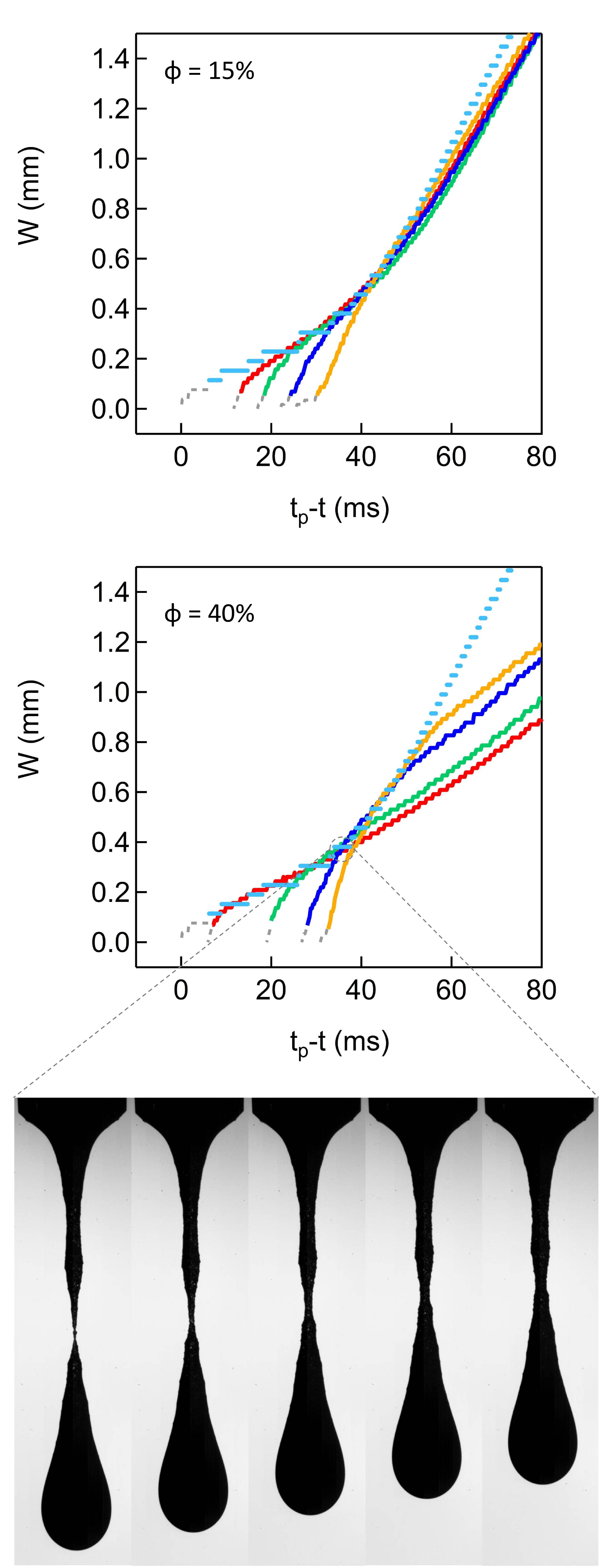}
                \caption[A set of four subfigures.]{Thinning dynamics for suspensions close to the deviation from the interstitial fluid regime for particle diameters of 20 $µm$ (\red{red}), 40 $µm$ (\green{green}), 80 $µm$ (\blue{blue}) and 140 $µm$ (\orange{orange}) for a volume fraction of: $\phi=15\%$ and $\phi=40\%$. The origin of the x-axis is given by the time of the pinch for the interstitial fluid, suspension curves have been shifted to match the interstitial fluid regime ($\phi=15\%$: $\Delta t_{20}=11.75$ ms, $\Delta t_{40}=17$ ms, $\Delta t_{80}=22$ ms \& $\Delta t_{140}=25.5$ ms. $\phi=40\%$: $\Delta t_{20}=5$ ms, $\Delta t_{40}=19$ ms, $\Delta t_{80}=26.75$ ms \& $\Delta t_{140}=31$ ms ). \textit{(bottom)} Corresponding snapshots for suspensions of $d=80 \mathrm{µm}$ particles  for $\phi = 40\%$ around the transition,  $\Delta t = \{-5, -2.5, 0, 2.5, 5 \}$ ms}.
                \label{fig:SE}%
                \end{figure}

Analogous to the study of the transition from the effective to the interstitial fluid regime we have studied the transition from the interstitial fluid regime to the final detachment regime. Figure \ref{fig:SE} \textit{(top)} shows the results for suspensions of different grain diameters of volume fraction $\phi=15\%$ and $\phi=40\%$ together with the interstitial fluid Shin Etsu. From these figures one can observe that there is a regime where the thinning dynamics are indeed identical for the suspensions and the interstitial fluid. One can also note that for the different grain diameters this regime can occur at very different neck radii and the thinning of the suspension and the interstitial fluid regime can be identical either during the exponential decay or during the linear regime corresponding to the Papageourgiou regime (see section \ref{sec:Theory}). From these figures it is also obvious that the final detachment of the suspensions is accelerated compared to the pure interstitial fluid.

\begin{figure}[h]
     \centering
   \includegraphics[width=0.45\textwidth]{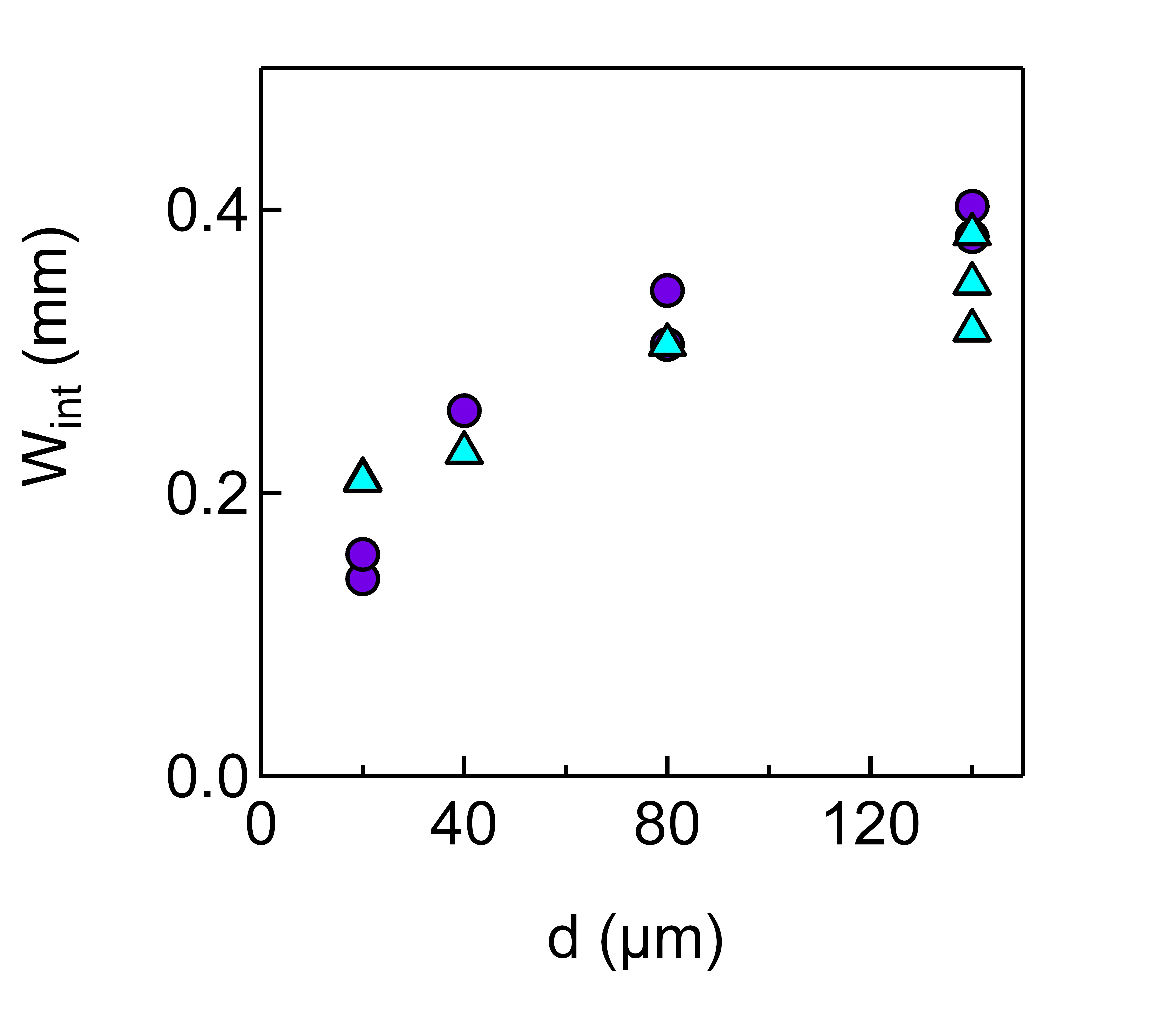}
\caption{Diameter of the neck at the transition between the interstitial fluid and the final detachment regime as a function of grain size for $\phi$: 15\% (\cyan{cyan}) and 40\% (\purple{purple})}
    \label{fig:W SE d}
\end{figure}

We have measured the neck radius $W_{int}$ at which the thinning dynamics start to differ between the suspensions and the interstitial fluid. On figure \ref{fig:W SE d} $W_{int}$ is represented as a function of the grain diameter $d$. For larger grains the crossover to the final detachment regime takes place earlier in the detachment process compared to smaller grains. Typically the crossover to the accelerated detachment regime takes place at a neck diameter that corresponds to some grain diameters. Once again the dependence of $W_{int}$ on $d$ is weak and the increase of $W_{int}$ with $d$ is observed to be less than linear. There might be a slight dependence on the grain fraction.

\subsection{\label{subsection:discussion} Discussion}

In the first two paragraphs of this section (section \ref{subsec:crossover_effective} and section \ref{subsec:crossover_SE}) we have shown that the three detachment regimes introduced in section \ref{sec:detachment regimes} are identical for different volume fractions and grain diameters. The crossover between the different regimes however is function of the volume fraction and the grain diameter. In this paragraph we discuss the visual observations from figures \ref{fig:regimes}, \ref{fig:PO} and \ref{fig:SE} and give possible interpretations of the origin of the mechanisms responsible for the different detachment regimes.

The snapshots in figure \ref{fig:PO} show images taken around the transition between the effective fluid regime and the interstitial fluid regime for a volume fraction of $\phi=40\%$ and $d=80\mcm$. First of all, one does not observe an obvious change in the visual appearance of the drops close to the transition. It is also clear that grains are still present in the neck at the transition. One observes that the drop seems to get rougher close to the transition to the interstitial fluid regime. Note that this roughness is not on the scale of individual particles but corresponds rather to aggregates of several particles. Royer \al \cite{Royer2009} have recently observed similar formation of aggregates during the destabilization of a dry granular jet. In our case rearrangements of the particles locally free space of grains in the neck of the suspension. This might be at the origin of the fact that the detachment process becomes independent of the volume fraction and is identical to the interstitial fluid, even if visual observations reveal that grains are still present in the neck.

In figure \ref{fig:SE} snapshots close to the crossover to the final detachment regime are shown. In this regime the neck diameter becomes of the order of magnitude of the grain size (few grain diameters). During the final detachment regime the deformation of the thread becomes very localized and the thinning does not seem to be a continuous process any more. The volume deformed in the neck is not continuous but decreases by steps. In this way the presence of grains prevents the formation of a long, stable filament.


\section{\label{sec:accelerated detachment}Acceleration}


\begin{figure}[h!]%
                \centering
                \subfigure[][ $\phi = 15 \%$]{%
                \label{fig:POF_Acceleration-a}%
                \includegraphics[width=7.5cm]{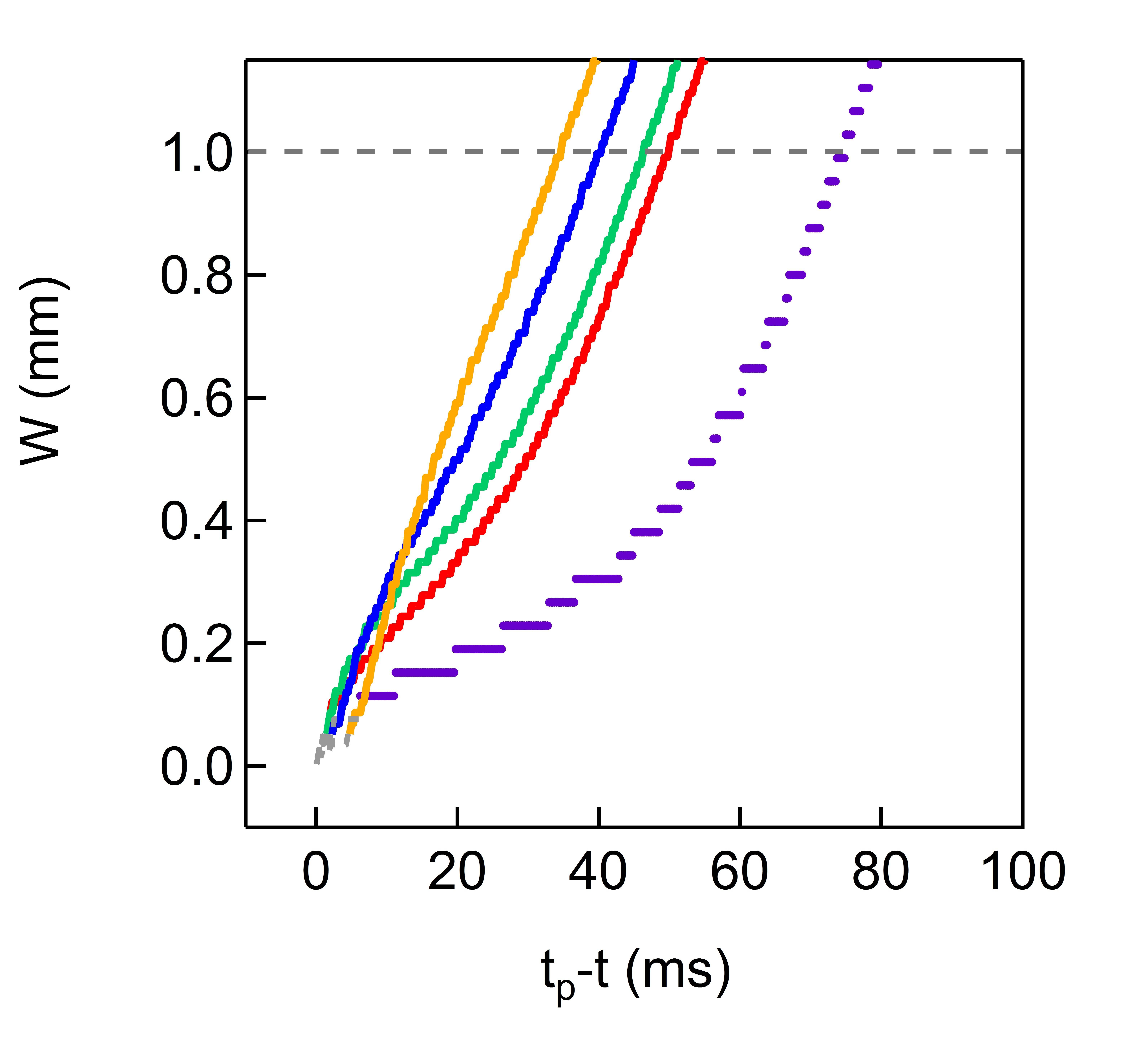}
                }%
              \\
                \subfigure[][ $\phi = 40 \%$]{%
                \label{fig:POF_Acceleration-b}%
                \includegraphics[width=7.5cm]{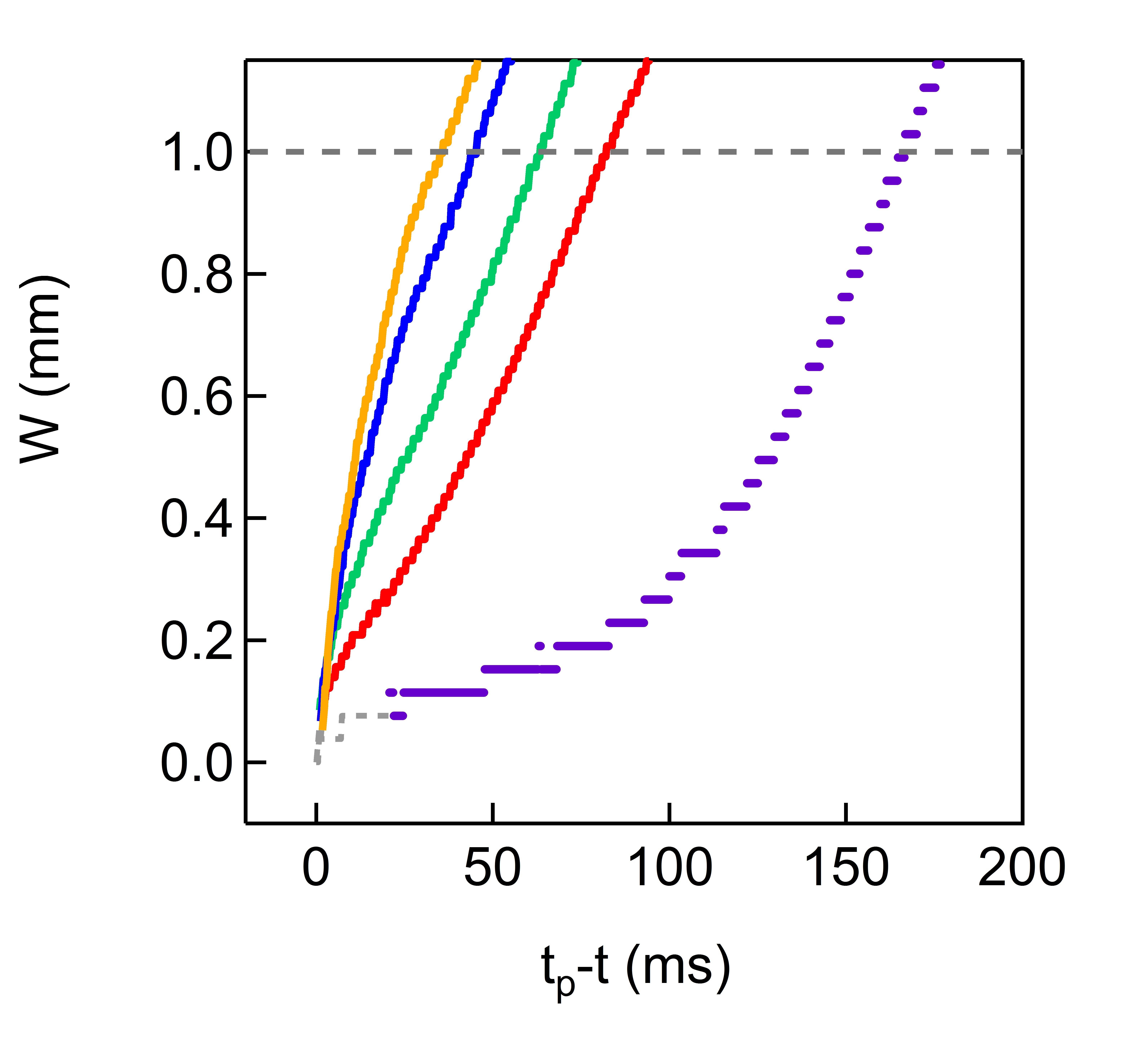}
                }
                \caption[A set of four subfigures.]{Characterization of the acceleration during the drop detachment process for different grain sizes: 20 $µm$ (\red{red}), 40 $µm$ (\green{green}), 80 $µm$ (\blue{blue}), 140 $µm$ (\orange{orange}) for volume fraction of: \subref{fig:POF_Acceleration-a} $\phi=15$ \%  and \subref{fig:POF_Acceleration-b} $\phi=40$ \%.}
                \label{fig:POF_Acceleration}
\end{figure}

Figure \ref{fig:POF_Acceleration} shows $W$ as a function of $t_p-t$ for $\phi=15\%$ and $\phi=40\%$ and different grain diameters together with the results for the pure oils AP200 and AP1000. All results are represented with respect to their real pinch off time. In this representation one observes that the suspensions detach significantly faster than the pure oils matching the shear viscosity of the suspensions. We have quantified this effect by measuring the rupture time $T_{rup}$ from the moment the minimal neck diameter reaches $W=1$~mm to the final pinch off for all suspensions and pure oils. Note that at $W=1$~mm all suspensions are in the effective fluid regime and their dynamics are at this stage identical to the dynamics of the pure oils. $T_{rup}$ is thus a good measure to quantify the acceleration due to the presence of particles in the thread.

\begin{figure}[h!]%
                \centering
                \subfigure[][ $\phi = 15 \%$]{%
                \label{fig:POF_DT-a}%
                \includegraphics[width=7.5cm]{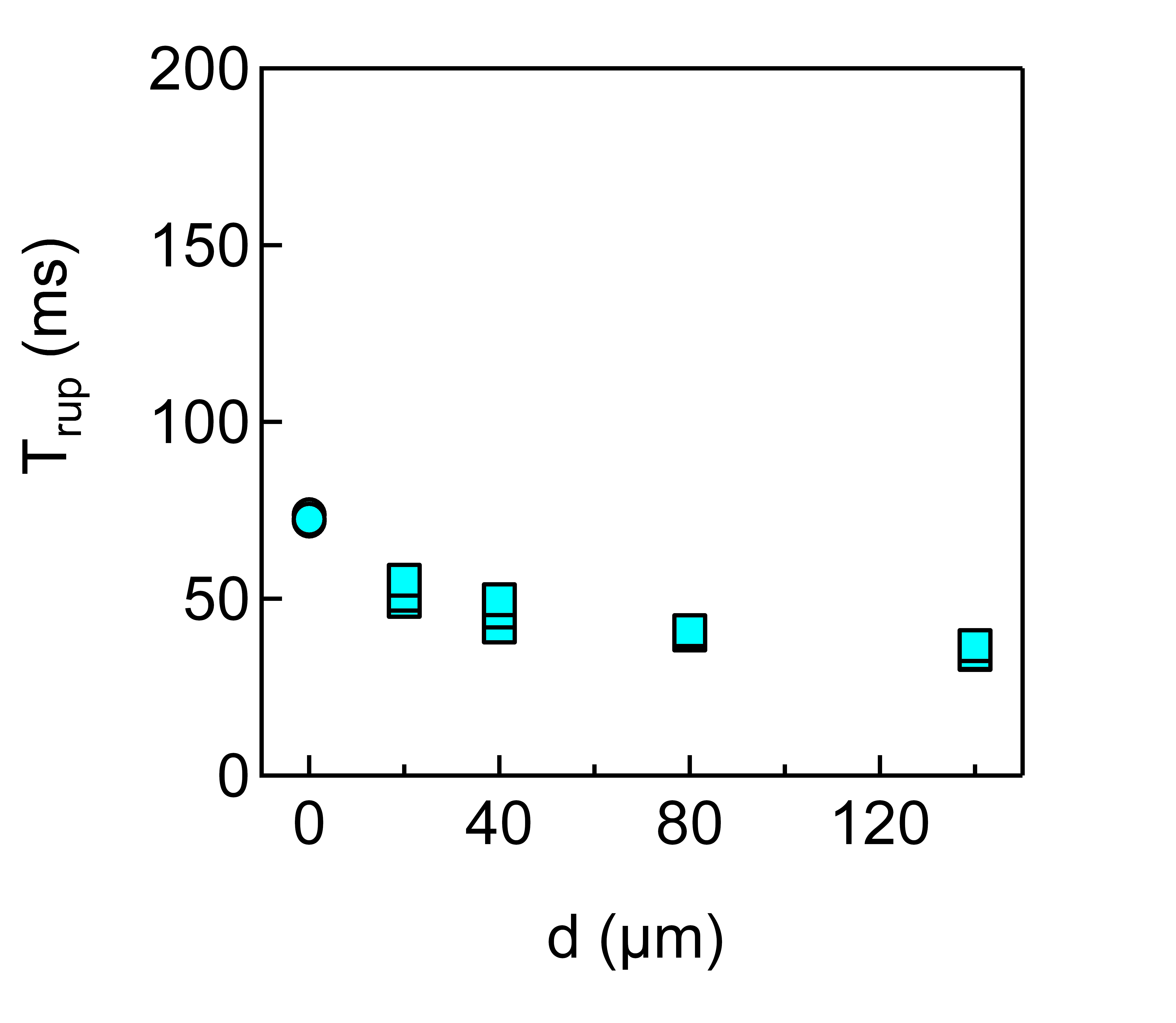}
                }%
              \\
                \subfigure[][ $\phi = 40 \%$]{%
                \label{fig:POF_DT-b}%
                \includegraphics[width=7.5cm]{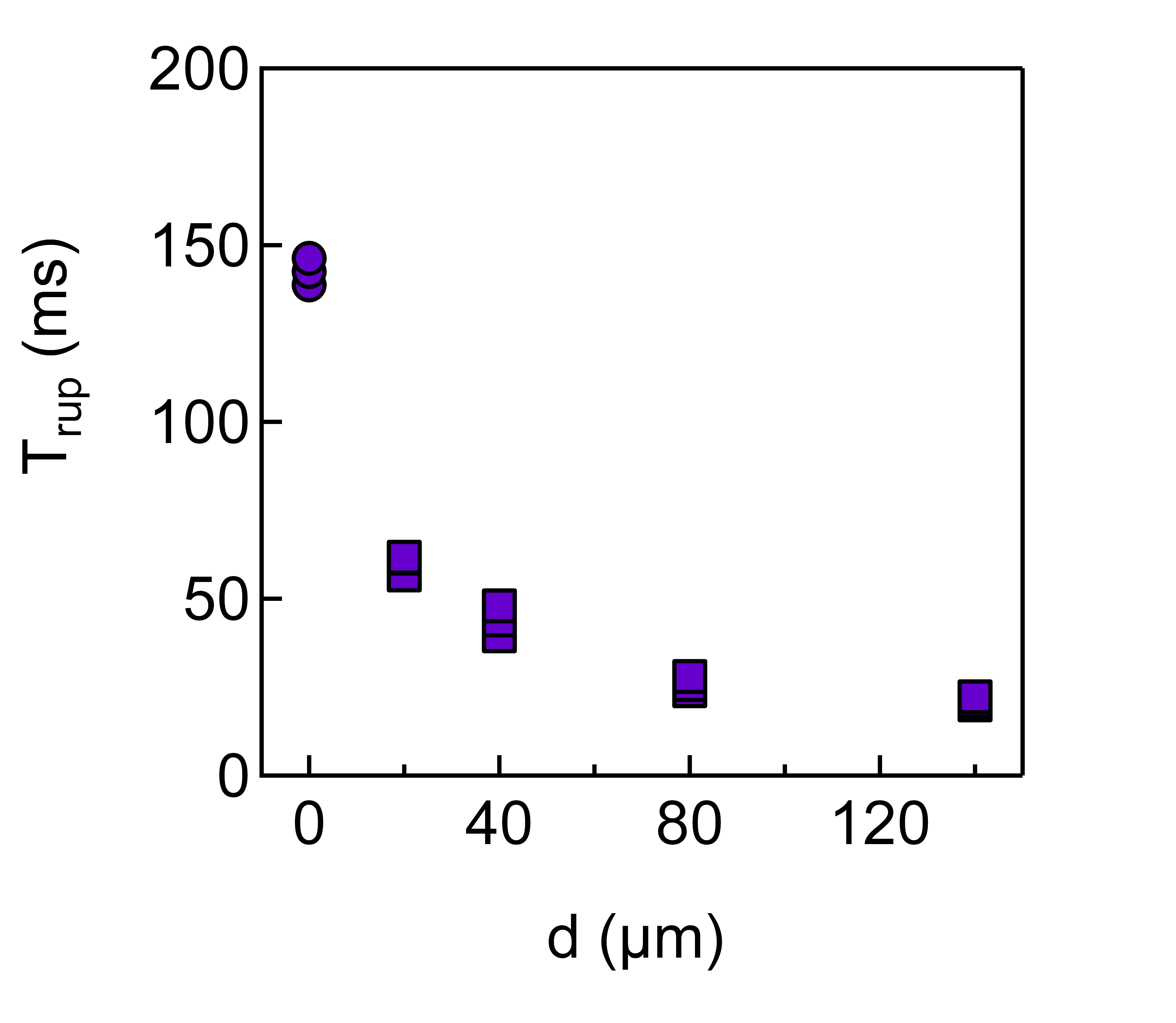}
                }
                \caption{Rupture time as a function of the particle diameter for  \subref{fig:POF_DT-a} $\phi=15$ \%  and \subref{fig:POF_DT-b} $\phi=40$ \%.}
    \label{fig:Dt_d}
\end{figure}

Figure \ref{fig:Dt_d} shows $T_{rup}$ as a function of the grain diameter $d$ for the two different volume fractions. In both cases one observes that the rupture time is function of the grain diameter $d$. The difference between the pure oil and the suspensions is more pronounced for the higher volume fraction than for the smaller volume fraction. This is due to the fact that the overall acceleration has two origins. First in the interstitial fluid regime the dynamics are given by the viscosity of the interstitial fluid $\eta_0$. The fact that $\eta_0$ is smaller than the shear viscosity of the suspensions leads to an acceleration of the detachment process. This acceleration is more important for the higher volume fractions as the difference in viscosity is higher for the latter. Then, the final detachment is accelerated for the suspensions compared to the interstitial fluid. The transitions in between these two regimes are function of the grain diameter. The total acceleration is thus function of the volume fraction and the grain diameter.


\section{\label{sec:conclusion}Conclusion}

In this paper we have shown that the detachment of a
drop of a dense granular suspension takes place through a number of
different regimes.

In the beginning of the detachment the suspensions
behave as an effective fluid and the dynamics of the detachment
process are identical to those observed for a pure oil of the same
shear viscosity. This indicates that the elongational viscosity of a granular suspension is solely given by the shear viscosity, as it is the case for a Newtonian fluid. At a certain stage of the detachment the drop enters a
different detachment regime. The thinning dynamics in this regime
 are found to be independent of the initial volume fraction and is identical to the dynamics of the pure interstitial fluid. The dynamics of the detachment of the suspensions thus correspond to the dynamics of a fluid without grains. Visual observations reveal however that grains are still present in the neck at this stage. We suspect rearrangements of grains to lead to density fluctuations in the neck, that becomes locally free of grains. These regions have a significantly lower viscosity compared to the suspensions and they thus dominate the thinning dynamics. The deviation from the effective fluid regime is function of the grain diameter and the volume fraction. Higher volume fractions and higher grain diameter lead to a deviation from the effective fluid regime at earlier stages of the detachment process, and thus at a larger neck diameter. This neck diameter is found to be of the order of 10 grain diameters, but is not linearly proportional to the latter.

The interstitial fluid regime is followed by a final detachment regime for the suspensions. The presence of grains in the neck prevents the formation of a stable and long filament as is observed for pure interstitial fluids. At the crossover to this regime, the thinning becomes very localized and the thinning of the filament does  not seem to be a continuous process any more. In other words the volume that is deformed is reduced. The deviation from the interstitial fluid regime is given by the grain diameter. It occurs typically at a neck radius of about 3 grain diameters, but it is once again not linearly proportional to the latter.

The fact that the detachment of a granular suspension takes place via these three different regimes leads to an overall acceleration of the detachment process compared to a pure oil matching the shear viscosity of the suspensions. This acceleration is function of the volume fraction and the grain diameter. It is more important for the higher volume fractions and the higher grain diameters. For larger grain diameters the crossover to the interstitial fluid regime and the accelerated regime takes place at earlier stages of the detachment, leading to a stronger acceleration. For the higher volume fractions the difference in viscosity between the interstitial fluid and the suspension is larger, leading also to a stronger difference between the detachment dynamics between the effective fluid and the interstitial fluid regime.

In summary, we have shown that the detachment of granular suspensions takes place via three different detachment regimes. We have characterized the crossover between these different regimes as a function of the volume fraction and the grain diameter. The detachment of the suspensions is accelerated compared to a pure fluid matching the shear viscosity of the suspensions. This acceleration is also function of the volume fraction and the grain diameter.

\begin{acknowledgments}
We acknowledge interesting discussions with Jens
Eggers, Christian Clasen, Daniel Bonn, Hamid Kellay, Alexander Morozov, Mark Miskin and
Heinrich Jaeger.
\end{acknowledgments}

\bibliographystyle{plain}

\end{document}